\documentclass[
  reprint,
  amsmath,amssymb, aps,
  prb,
]{revtex4-1}

\usepackage{graphicx}
\usepackage{dcolumn}
\usepackage{hyperref}
\usepackage{cancel} 
\usepackage[normalem]{ulem}
\usepackage[usenames, dvipsnames]{color}
\usepackage[normalem]{ulem}
\usepackage{bm}
\usepackage{booktabs}

\begin{document}
\newcommand{\kb}{$\mathbf{k}$}

\preprint{APS/123-QED}

\title{Generalized Regular \kb-point Grid Generation On The Fly}

\author{Wiley S. Morgan, John E. Christensen, Parker K. Hamilton, Jeremy
  J. Jorgensen, Branton J. Campbell, Gus L. W. Hart}
\affiliation{Department of Physics and Astronomy, Brigham Young
  University, Provo, Utah, 84602, USA}

\author{Rodney W. Forcade} \affiliation{Department of Mathematics,
  Brigham Young University, Provo, Utah, 84602, USA}


\date{\today}
\begin{abstract}
In the DFT community, it is common practice to use \textit{regular}
\kb-point grids (Monkhorst-Pack, MP) for Brillioun zone
integration. Recently Wisesa et. al.\cite{wisesa2016efficient} and
Morgan et. al.\cite{MORGAN2018424} demonstrated that
\textit{generalized regular} (GR) grids offer advantages over
traditional MP grids. GR grids have not been widely adopted because
one must search through a large number of candidate grids. This work
describes an algorithm that can quickly search over GR grids for those
that have the most uniform distribution of points and the best
symmetry reduction. The grids are $\sim$60\% more efficient, on
average, than MP grids and can now be generated on the fly in seconds.
\end{abstract}

\maketitle






\section{Introduction}
In computational materials science, the properties of crystalline
materials are often calculated using density functional theory
(DFT). These codes integrate the electronic energy over occupied
states in the Brillouin zone. In the case of metals, convergence is
very slow. The convergence rate is proportional to the density of
\kb-points used to sample the Brillouin zone. An order of magnitude
increase in accuracy an order of magnitude more \kb-points.

Additionally, as high throughput\cite{curtarolo2012aflow,
  saal2013materials, jain2013commentary, NOMAD,
  landis2012computational, hachmann2011harvard, hummelshoj2012catapp,
  de2015database, de2015charting, cheng2015accelerating,
  gomez2016design, chan2015combinatorial, tada2014high,
  pilania2013accelerating, yan2015material, ramakrishnan2014quantum,
  hachmann2014lead, lin2012silico, armiento2014high,
  senkov2015accelerated} calculations have become more popular because
of their recent successes \cite{greeley2006computational,
  gautier2015prediction, oliynyk2017discovery,
  chen2012carbonophosphates, hautier2011phosphates, jahne2013new,
  moot2016material, aydemir2016ycute, zhu2015computational,
  chen2016understanding, ceder1998identification, yan2015design,
  bende2017chemical, mannodi2017scoping, sanvito2017accelerated,
  yaghoobnejad2016combined, hautier2013identification, bhatia2015high,
  ceder1998identification, johannesson2002combined,
  stucke2003predictions, curtarolo2005accuracy, matar2009first,
  ceder2011recharging, sokolov2011computational, ulissi2017machine,
  levy2009new, ma2013improved, yang2012search, chen2012synthesis,
  kirklin2013high}, the accuracy of the calculations becomes more
important. The accuracy and quantity of calculations within material
databases is a crucial component in high throughput and machine
learning approaches. Increasing the speed of calculations, without
reducing the accuracy, would significantly impact material
predictions.

DFT codes generally use regular grids, proposed by Monkhorst and Pack
(MP)\cite{monkhorst1976special}, to define their \kb-point
grids. \kb-points within a regular grid are defined by:
\begin{equation}
  \begin{split}
    \bold{k} & = (\bold{b}_1, \bold{b}_2, \bold{b}_3) \mathbb{D}^{-1}\begin{pmatrix} n_1 \\ n_2 \\ n_3 \end{pmatrix} \\
    & = \frac{n_1}{d_1} \bold{b}_1+\frac{n_2}{d_2} \bold{b}_2+\frac{n_3}{d_3} \bold{b}_3
  \end{split}
\end{equation}
where $\bold{b}_i$ are the reciprocal lattice vectors, $\mathbb{D}$ is
a diagonal integer matrix with $d_i$ along the diagonal, and $n_i$
runs from 0 to $d_{i-1}$.

An alternative, more general method was proposed by Moreno and
Soler,\cite{moreno1992optimal} which involves searching through grids
at a desired \kb-point density for those that have the highest
symmetry reduction, i.e., the lowest general-point multiplicity or
fewest symmetrically distinct \kb-points. High symmetry reduction
impacts the computations cost, the cost of a DFT calculation scales
with the number of irreducible \kb-points. The grids are then sorted
by the length of the shortest grid generating vector and the grid with
the longest vector is choosen, thus selecting the most uniform
grid. The Moreno-Soler method involves the construction of
superlattices from the real-space parent lattice (primitive lattice)
\begin{equation}
  (\bold{s}_1, \bold{s}_2, \bold{s}_3) = (\bold{a}_1, \bold{a}_2, \bold{a}_3) \mathbb{H}
\end{equation}
where the columns $\bold{s}_i$ are the supercell vectors, the columns
$\bold{a}_i$ are the parent lattice vectors, and $\mathbb{H}$ is an
integer matrix. The \emph{dual} lattice of the superlattice vectors
supercell lattice then defines a set of \kb-point grid generating
vectors $\bm{\kappa}_i$.
\begin{equation}
  \begin{split}
    (\bm{\kappa}_1, \bm{\kappa}_2, \bm{\kappa}_3) & = 2 \pi ((\bold{s}_1, \bold{s}_2, \bold{s}_3)^{-1})^{T} \\
    & = 2 \pi (((\bold{a}_1, \bold{a}_2, \bold{a}_3) \mathbb{H})^{-1})^{T} \\
    & = 2 \pi (\mathbb{H}^{-1})^{T} ((\bold{a}_1, \bold{a}_2, \bold{a}_3)^{-1})^{T} \\
    & = (\mathbb{H}^{-1})^{T} (\bold{b}_1, \bold{b}_2, \bold{b}_3) 
  \label{eq:dual}
  \end{split}
\end{equation}
Note that the determinant of $\mathbb{H}$ determines the number of
\kb-points that lie within the Brillouin zone.

We refer to grids generated by the Moreno-Soler method as
\emph{Generalized Regular} (GR) grids. GR grids have never been widely
adopted because they require a search over many supercells to select
the cell that 1) maximizes the distance between points and 2) have the
fewest irreducible \kb-points, i.e., has the highest symmetry
reduction. These searches tend to be time consuming due to the
combinatoric explosion in the total number of possible supercells
shown in Fig.~\ref{fig:Num_HNFs}.

Recently Wisesa, McGill, and Mueller\cite{wisesa2016efficient} (WMM)
rectified this by creating a \kb-point server containing precalculated
grids that have high symmetry reduction. These grids can be retrieved
via an internet request and have been demonstrated to be 60\% more
efficient than MP grids \cite{MORGAN2018424}. However, the requirement
of an internet query, which cannot be performed in typical
supercomputer environments, makes them difficult to use in some
cases. Here we present an algorithm for generating GR grids ``on the
fly'' (avoiding the need for an internet query). This algorithm has
been implemented in a code available at
https://github.com/msg-byu/GRkgridgen. This code takes the numerical
lattice vectors, atomic basis vectors, and grid density from a user
and returns the optimal GR grid.

\section{Methodology}

\subsection{Generating Symmetry-Preserving Supercells}

\begin{figure}[t]
  \centering
  \includegraphics[width=9cm]{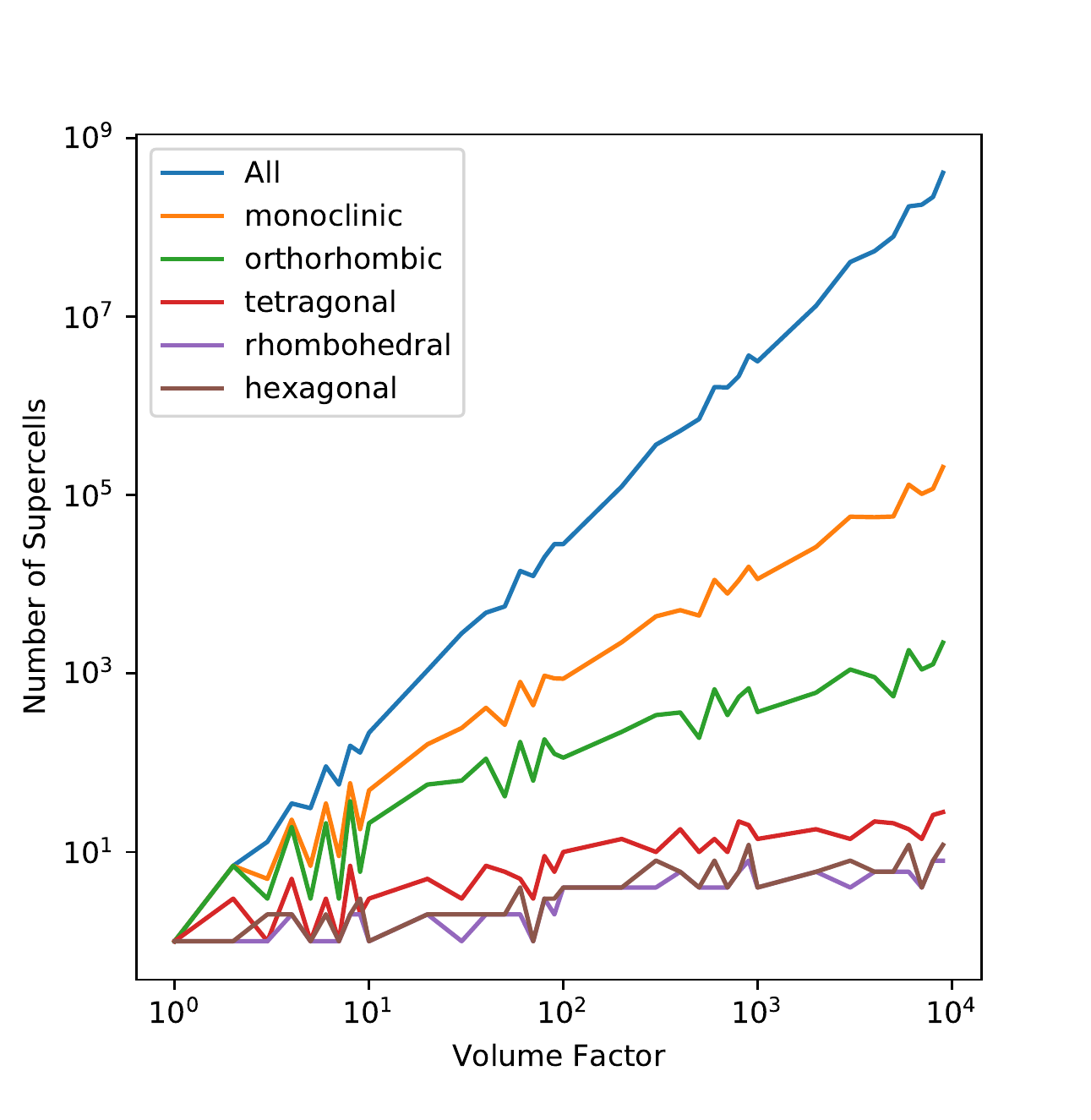}
  \caption{The number of supercells that preserve the symmetry of the
    parent cell at various volume factors. The total number of
    supercells that exist is also displayed for comparison. Cubic
    cells were omitted since they have at most one symmetry-preserving
    supercell at an given volume factor.}
  \label{fig:Num_HNFs}
\end{figure}

The main difficulty in generating GR grids is that the number of
distinct supercells grows rapidly with the volume factor (the
determinant of $\mathbb{H}$). \footnote{Note that the determinant of
  $\mathbb{H}$ determines the number of \kb-points in the Brillouin
  zone.} To optimize the \kb-point folding efficiency, the \kb-point
grid should have the same symmetry as the parent cell. The number of
supercells that preserve the symmetry of the parent is always
significantly smaller than the number of possible supercells (except
in the case of triclinic lattices) as can be seen in
Fig.~\ref{fig:Num_HNFs}. If one can quickly generate only those
supercells that preserve the symmetry of the parent, avoiding the
combinatorial explosion, the computational burden is drastically
reduced.

To generate only the symmetry-preserving supercells, we restrict
$\mathbb{H}$ to be an integer matrix in Hermite Normal Form (HNF)
subject to the constraints:
\begin{equation}
  \begin{split}
    & \mathbb{H} = \begin{pmatrix} a \; 0 \; 0\\ b \; c \; 0 \\ d \; e \; f \end{pmatrix} \\
    & a, \, c, \, f > 0 \\
    & b \geq 0, \; \; \; b < c \\
    & d, e \geq 0, \; \; \; d, e < f 
  \end{split}
  \label{eq:hnf}
\end{equation}
We will use the notation that $\mathbb{A} = (\bold{a}_1, \bold{a}_2,
\bold{a}_3)$ is the parent lattice and $\mathbb{C} = (\bold{c}_1,
\bold{c}_2, \bold{c}_3)$ is a supercell such that $\mathbb{C} =
\mathbb{A} \mathbb{H}$. When the lattice symmetries are applied to
$\mathbb{A}$, they generate another set of basis vectors $\mathbb{A}'$
\begin{equation}
  \mathbb{A}' = \bold{g}\mathbb{A}
\end{equation}
(where $\bold{g}$ is an element of the point group). Because
$\mathbb{A}$ and $\mathbb{A'}$ are related by a symmetry operation of
the lattice, they both represent the same lattice and are related by
an integer matrix
\begin{equation}
  \begin{split}
    \mathbb{A}' = \mathbb{A}\mathbb{X} \\
    \mathbb{A}\mathbb{X} = \bold{g}\mathbb{A} \\
    \mathbb{X} = \mathbb{A}^{-1}\bold{g}\mathbb{A} \\
  \end{split}
  \label{eq:int_ops}
\end{equation}
where $\mathbb{X}$ is an integer matrix with determinant
$\pm{1}$. Similarly, if a supercell $\mathbb{C}$ has the same symmetry
as $\mathbb{A}$ then all the symmeties of $\mathbb{A}$ will map
$\mathbb{C}$ to another basis $\mathbb{C}'$ that will be related to
$\mathbb{C}$ by a unimodular transformation
\begin{equation}
  \begin{split}
    & \mathbb{C}' = \bold{g} \mathbb{C} \; \forall \; \bold{g} \in \bold{G} \\
    & \mathbb{C} \mathbb{M} = \bold{g} \mathbb{C} \\ 
    & \mathbb{M} = \mathbb{C}^{-1} \bold{g} \mathbb{C}
  \end{split}
  \label{eq:int_origins}
\end{equation}
where $\bold{G}$ is the set of generators of the point group of
$\mathbb{A}$ and $\mathbb{M}$ is an integer matrix. Using
Eqs. (\ref{eq:int_ops}) and (\ref{eq:int_origins}), it is possible to
define restrictions on the entries of $\mathbb{H}$:

\begin{equation}
  \mathbb{M} = \mathbb{H}^{-1}\mathbb{X}\mathbb{H}.
  \label{eq:int_rels}
\end{equation}

In other words $\mathbb{H}$ must be such that $\mathbb{M}$ is
transformation of $\mathbb{X}$ that retains integer entries. Equation
(\ref{eq:int_rels}) yields the following system of linear equations
\begin{equation}
  \begin{split}
    & \alpha_1 = \frac{bx_{12}+dx_{13}}{a} \\
    & \alpha_2 = \frac{cx_{12}+ex_{13}}{a} \\
    & \alpha_3 = \frac{fx_{13}}{a} \\
    & \beta_1 = \frac{-bx_{11}+ax_{21}-b\alpha_1+bx_{22}+dx_{23}}{c} \\
    & \beta_2 = \frac{-b\alpha_2+ex_{23}}{c} \\
    & \beta_3 = \frac{-b\alpha_3+cx_{23}}{c} \\
    & f = \frac{\alpha_4}{c} \\
    & \gamma_1 = \frac{ax_{31}+bx_{32}+dx_{33}-e\beta_1-d\alpha_1-dx_{11}}{f} \\
    & \gamma_2 = \frac{-ex_{22}+cx_{32}+ex_{33}-e\beta_2-d\alpha_2}{f} \\
    & n = a \cdot c \cdot f 
  \end{split}
  \label{eq:all_rels}
\end{equation}
where $x_i$ are the entries of $\mathbb{X}$, $n$ is the determinant of
$\mathbb{H}$ and $\alpha_i$, $\beta_i$, and $\gamma_i$ are arbitrary
names for the expressions used for convenience. $\mathbb{H}$ will
generate a supercell that preserves the symmetries of $\mathbb{A}$
when $\alpha_1$, $\alpha_2$, $\alpha_3$, $\alpha_4$, $\beta_1$,
$\beta_2$, $\beta_3$, $\gamma_1$, and $\gamma_2$ are all integers for
each generator in $\bold{G}$. Even though the solutions to
(\ref{eq:all_rels}) have no closed form, we may use them to build an
algorithm that generates $\mathbb{H}$ matrices that preserve the
lattice symmetries.

The specific form of $\mathbb{X}$ depends on the basis chosen for the
parent lattice, the solutions to (\ref{eq:all_rels}), and resulting
algorithms, will differ depending on the basis. For example, if a
base-centered orthorhombic lattice is constructed with the basis
\begin{equation}
  \mathbb{A}_1 = (\bold{a}_1, \bold{a}_2, \bold{a}_3) = \begin{pmatrix} \begin{matrix} \frac{1}{2} \\ 1 \\ 0 \end{matrix} & \begin{matrix} \frac{1}{2} \\ -1 \\ 0 \end{matrix} & \begin{matrix} 0 \\ 0 \\ 3 \end{matrix} \end{pmatrix}
\end{equation}
then (\ref{eq:all_rels}) would reduce to (each equation has three
outputs because the base centered orthormbic point-group has three
generators):
\begin{equation}
  \begin{split}
    & \alpha_1 = \begin{pmatrix} 0, \; 0, \; -\frac{b}{a} \end{pmatrix} \\
    & \alpha_2 = \begin{pmatrix} 0, \; 0, \; -\frac{c}{a} \end{pmatrix} \\
    & \alpha_3 = \beta_3 = \begin{pmatrix} 0, \; 0, \; 0 \end{pmatrix} \\
    & \beta_1 = \begin{pmatrix} 0, \; 0, \; \frac{-a - b\alpha_1}{c} \end{pmatrix} \\
    & \beta_2 = \begin{pmatrix} 0, \; 0, \; \frac{b}{a} \end{pmatrix} \\
    & \gamma_1 = \begin{pmatrix} 0, \; \frac{2d}{f}, \; \frac{-d-d\alpha_1-e\beta_1}{f} \end{pmatrix} \\
    & \gamma_2 = \begin{pmatrix} 0, \; \frac{2e}{f}, \; \frac{-e-d\alpha_2-e\beta_2}{f} \end{pmatrix} \\
  \end{split}
  \label{eq:ex_rels}
\end{equation}
All the equations in (\ref{eq:ex_rels}) must be simultaneously
satisfied for the generated $\mathbb{H}$'s to preserve the symmetries
of $\mathbb{A}_1$. Alternatively the basis
\begin{equation}
  \mathbb{A}_2 = (\bold{a}_1, \bold{a}_2, \bold{a}_3) = \begin{pmatrix} \begin{matrix} \frac{1}{2} \\ 1 \\ 0 \end{matrix} & \begin{matrix} 0 \\ -2 \\ 0 \end{matrix} & \begin{matrix} 0 \\ 0 \\ 3 \end{matrix} \end{pmatrix}
\end{equation}
could be used to construct the same lattice. When basis $\mathbb{A}_2$
is chosen, the relations in (\ref{eq:all_rels}) become:
\begin{equation}
  \begin{split}
    & \alpha_1 = \alpha_2 = \alpha_3 = \beta_2 = \beta_3 = \begin{pmatrix} 0, \; 0, \; 0 \end{pmatrix} \\
    & \beta_1 = \begin{pmatrix} 0, \; 0, \; \frac{a + 2b}{c} \end{pmatrix} \\
    & \gamma_1 = \begin{pmatrix} 0, \; \frac{2d}{f}, \; \frac{-e\beta_1}{f} \end{pmatrix} \\
    & \gamma_2 = \begin{pmatrix} 0, \; \frac{2e}{f}, \; -\frac{2e}{f} \end{pmatrix} \\
  \end{split}
\end{equation}

Note the stark difference between the relationships derived from
$\mathbb{A}_1$ and $\mathbb{A}_2$. $\mathbb{A}_2$ results in fewer
equations to check, however, $\mathbb{A}_1$ gives relationships
between $a$ and $b$, and $a$ and $c$ separately resulting in a faster
search since many combinations can be skipped early in the search.  By
taking care in selecting a basis for each lattice, one can find an
efficient set of conditions for generating the supercells of that
basis.

\subsection{Niggli Reduction}

Choosing a basis for each type of lattice presents a problem; there
are an infinite number of lattices basis choices. The number of bases
is substantially reduced by recognizing that any given
symmetry-preserving HNF, $\mathbb{H}^{\textrm{sp}}$, will work for
\textit{every} lattice of the same symmetry. The sensitivity of the
representation of the point group $\mathbb{X}$ on the chosen basis
requires a set of representative bases that goes beyond the 14 Bravais
lattices. Such a set was constructed by Niggli\cite{Kiiv1976,
  SantoroMighell, Grosse-Kunstleve:sh5006, SantoroMighell,
  ITCcite-key}, who identified 44 distinct bases. Any given basis of a
crystal can be classifed as one of these 44 cases by reducing it to
the Niggli canonical form and then comparing the lengths of the basis
vectors and the angles between them. If two nominally different
lattices reduce to the same Niggli case, then the two lattices are
``equivalent'' and have the same symmetries and the same set of
$\mathbb{H}^{\textrm{sp}}$s.

Niggli reduction allows for the user's basis to be mapped to a basis
which has convenient solutions to Eqs.~(\ref{eq:all_rels}). The
strategy is to define the $\mathbb{H}^{\textrm{sp}}$'s in the selected
basis, then generate the supercells for the selected basis and
transform them to the $\mathbb{H}$'s for the Niggli reduced basis,
$\mathbb{H}^{\textrm{sp}}_R$. Once the $\mathbb{H}^{\textrm{sp}}_R$'s
have been determined, they can be applied directly to the user's
reduced basis to create a symmetry-preserving supercell of the user's
parent cell and thus define an efficient \kb-point grid at the
specified density.

\subsection{Grid Selection}

\begin{figure*}
  \centering
  \includegraphics[width=\textwidth]{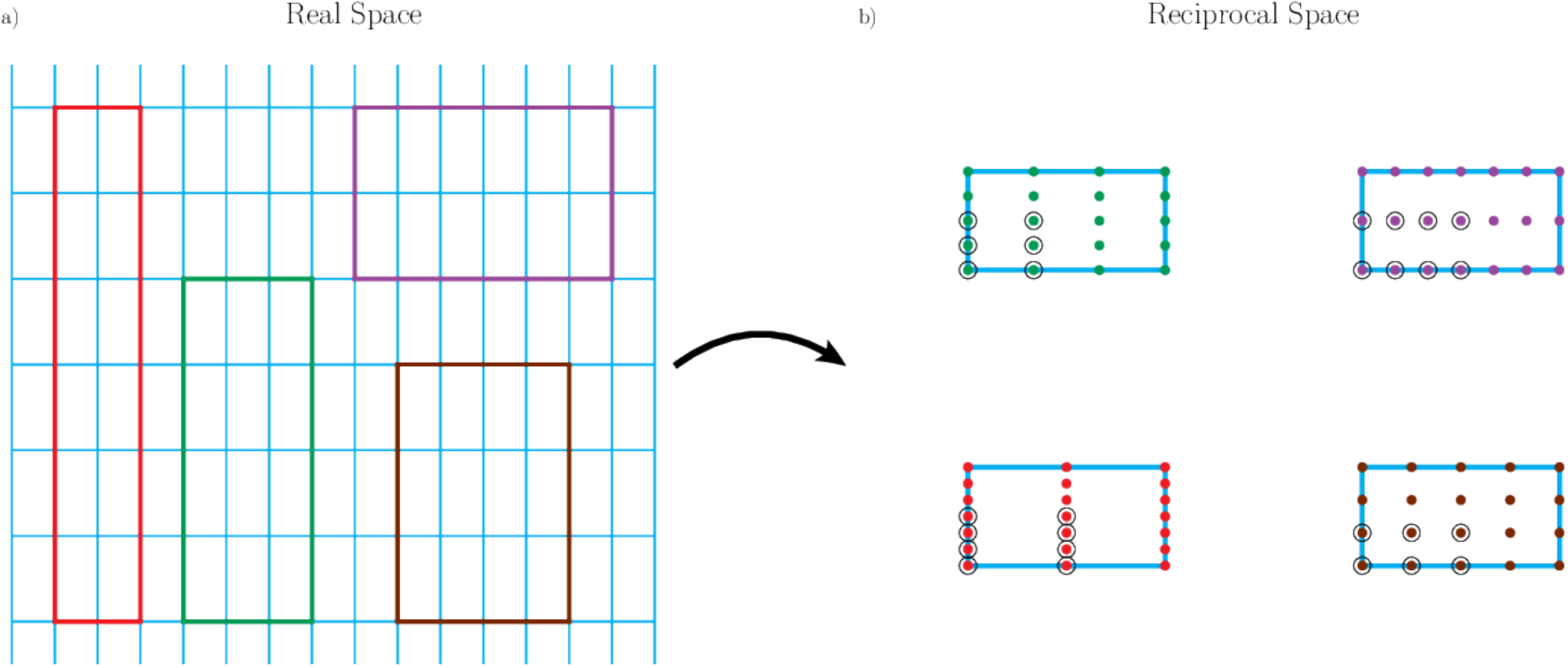}
  \caption{A 2D example of symmetry-preserving supercells and the
    \kb-point grids that they would generate for a rectangular
    lattice. a) contains four symmetry preserving supercells of the
    primitive cell, shown in blue, with a volume factor of 12. In b)
    the primitive cell, blue cells, and the supercells have been
    mapped to reciprocal space and the grids that would have been
    generated from each supercell have been placed in a cell. The
    color of the grid points matches the color of the generating
    supercell. The circled points are the irreducible \kb-points of
    each grid.}
  \label{fig:real_recip}
\end{figure*}

At a given volume factor (i.e., number of \kb-points), the integer
relations in Eq. (\ref{eq:all_rels}) will yield multiple supercells
for most lattices, a 2D example of these supercells is provided in
Fig.~\ref{fig:real_recip}(a). It is then neccessary to select one
which defines the best \kb-point grid. This is done by transforming
each symmetry-preserving supercell to its corresponding \kb-point grid
generating vectors as in Eq. \ref{eq:dual}; see
Fig.~\ref{fig:real_recip}(b). We then search this set of grids for one
that has optimal properties---a uniform distribution of points and the
best symmetry reduction. To ensure the grid generating vectors are as
short as possible we perform Minkowski reduction\cite{mink_red}, then
sort the grids by the length of their shortest vector.

The most uniform grids will have the maximal shortest vector. We
filter the grids so that none with a packing fraction of less then
$0.3$ are considered. Each of the uniform grids is then symmetry
reduced\cite{kpfolding} in order to determine which has the fewest
irreducible \kb-points. Table~\ref{tab:grid_props} shows the length of
the shortest vector and number of irreducible \kb-points for the grids
in Fig.~\ref{fig:real_recip}(b). The grids are sorted first by the
length of their shortest vector (eliminating the green and red grids)
then by the number of irreducible \kb-points such that the ideal grid
appears at the top of the table, i.e., the grid generated by the brown
supercell in Fig.~\ref{fig:real_recip}(a).

\begin{table}[t]
  \centering
  {\renewcommand{\arraystretch}{1.35}
  \begin{tabular}{|c|c|c|}
    \hline
    grid   & shortest vector length & number of irreducible \kb-points \\
    \hline
    brown  & $\frac{1}{6}$ & 6 \\
    \hline
    purple & $\frac{1}{6}$ & 8 \\
    \hline
    green  & $\frac{1}{8}$ & 6 \\
    \hline
    red    & $\frac{1}{12}$ & 8 \\
    \hline
  \end{tabular}}
  \caption{Properties (length of shortest vector and number of
    irreducible \kb-points) of the grids in Fig.~\ref{fig:real_recip}}
  \label{tab:grid_props}
\end{table}

\begin{figure*}
  \centering
  \includegraphics[width=\textwidth]{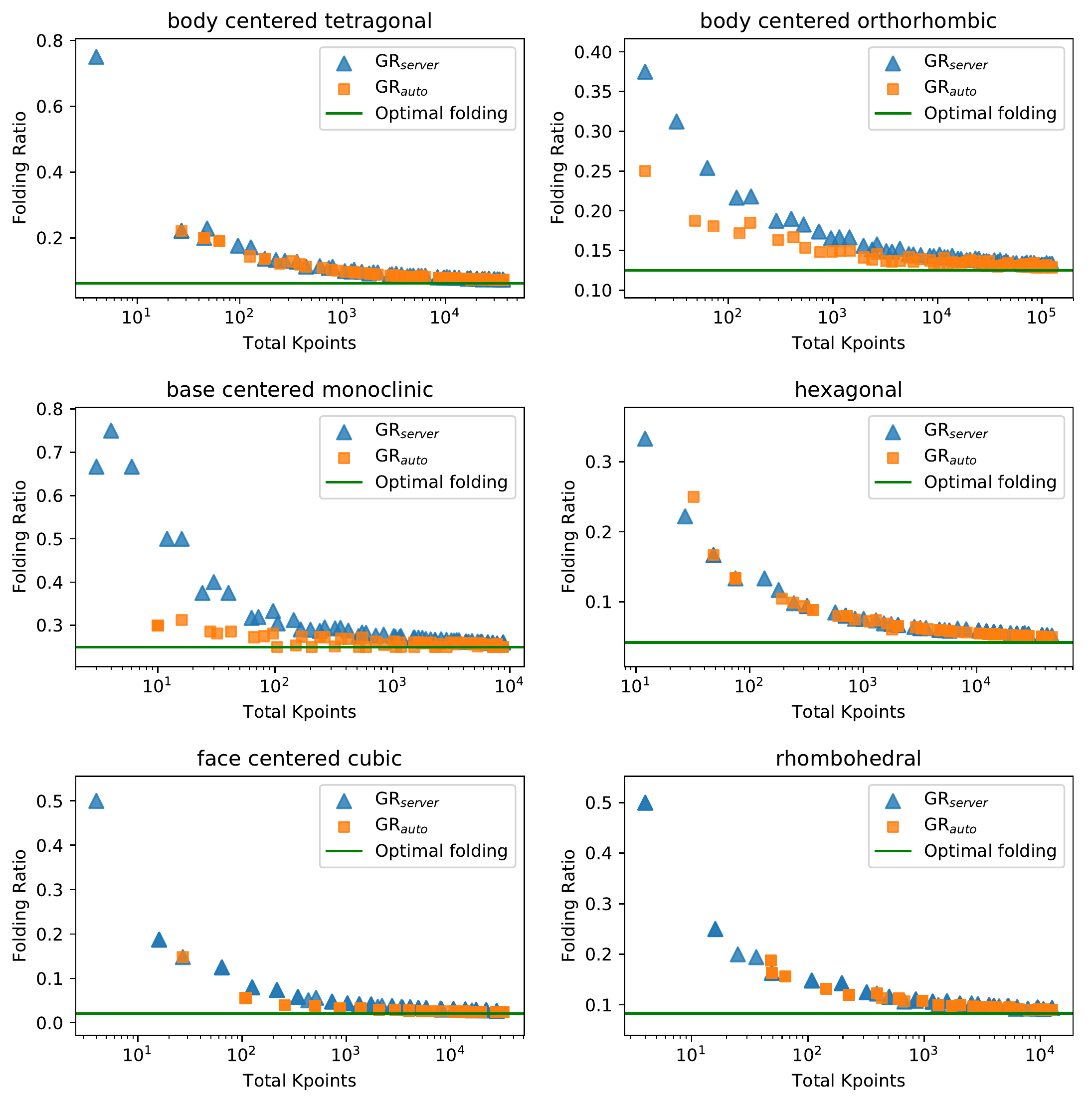}
  \caption{A comparison of the $GR_{\textrm{auto}}$ and
    $GR_{\textrm{server}}$ \kb-point grids. For each grid th number of
    irreducible \kb-points was divided by the total number of
    \kb-points. This shows that both sets of grids offer similar
    folding at a given \kb-point destiy and will have similar
    efficiencies.}
  \label{fig:ratio_compare}
\end{figure*}

It is also possible to offset the \kb-point grid from the origin to
improve the grids efficiency. The origin is not symmetrically
equivalent to any other point in the grid; for example, including an
offset makes it possible for the point at the origin to be mapped to
other points in the grid, decreasing the number of irreducible
\kb-points. Different grids have different symmetry-preserving offsets
that should be tested. For example, both simple cubic and
face-centered cubic (fcc) grids have one possible offset that
preserves the full symmetry of the cell, $(\frac{1}{2}, \frac{1}{2},
\frac{1}{2})$ (expressed as fractions of the grid generating vectors),
while a body-centered-cubic lattice has no symmetry preserving
offsets\footnote{For some lattices no symmetry preserving offsets
  exist. In these cases using an offset that does not preserve the
  full symmetry can be beneficial. For example, a body centered cubic
  system with an offset of $(0, 0, \frac{1}{2})$ can sometimes offer
  better folding than the same grid with no offset.}, and simple
tetragonal has three symmetry preserving offsets. (For a full list of
the symmetry-preserving offsets by lattice type, see the appendix.)
The grid that has the fewest \kb-points with a given offset is
selected.

Not every volume factor will have a symmetry-preserving grid that is
uniform. To ensure that a symmetry-preserving grid is found, it is
necessary to include multiple volume factors in the search. The number
of additional volume factors to search depends on the lattice type; in
general, the search should continue until multiple candidate grids
have been found. The best grid is then selected from these candidates.

\subsection{Method Summary}

The algorithm can be summarized in the following steps:

\begin{enumerate}
  \item Identify the Niggli reduced cell of the user's structure.
  \item Generate the symmetry-preserving HNFs for the canonical form
    of the Niggli cell.
  \item Map the resulting supercells to the original lattice using the
    Niggli-reduced basis as an intermediary.
  \item Convert the supercells into \kb-point grid generating vectors.
  \item Perform Minkowski reduction on the grid generating vectors.
  \item Sort the grid generating vectors by the length of their
    shortest vector.
  \item Select the grids that maximize the length of the shortest
    vectors.
  \item Use the symmetry group to reduce the selected grids to find
    the one with the fewest irreducible \kb-points.
\end{enumerate}
 
\section{Results}

To test the above algorithm, we compared the \kb-point grids it
generates, $GR_{\textrm{auto}}$, to those generated by the \kb-point
sever\cite{wisesa2016efficient}, $GR_{\textrm{server}}$ in two
ways. First, we generated both grids over a range of \kb-point
densities for over 100 crystal lattices. These lattices were
constructed for nine elemental systems---Al, Pd, Cu, W, V, K, Ti, Y,
and Re---with supercells for the cubic systems having between 1--11
atoms per cell and supercells for the hexagonal close packed systems
having between 2--14 atoms per cell. Additional test structures were
selected from AFLOW\cite{curtarolo2012aflow}. All tests were conducted
without offsetting the grids from origin. We then plotted the
resulting ratio of irreducible \kb-points to total \kb-points in each
grid. Six representative examples of the results are shown in
Fig.~\ref{fig:ratio_compare}. These tests show that the
$GR_{\textrm{auto}}$ grids should be very close in performance to
$GR_{\textrm{server}}$ grids. Additionally, the tests show that
convergence toward the ideal folding ratio is rapid for all lattice
types.


The second test compared the total energy errors of MP (generated by
AFLOW), $GR_{\textrm{auto}}$ and $GR_{\textrm{server}}$ grids in the
same manner, and using the same methods, as done in our previous study
of $GR$ grids\cite{MORGAN2018424}. We provide a brief review of that
method here.

\begin{figure}[t]
  \centering
  \includegraphics[width=8cm]{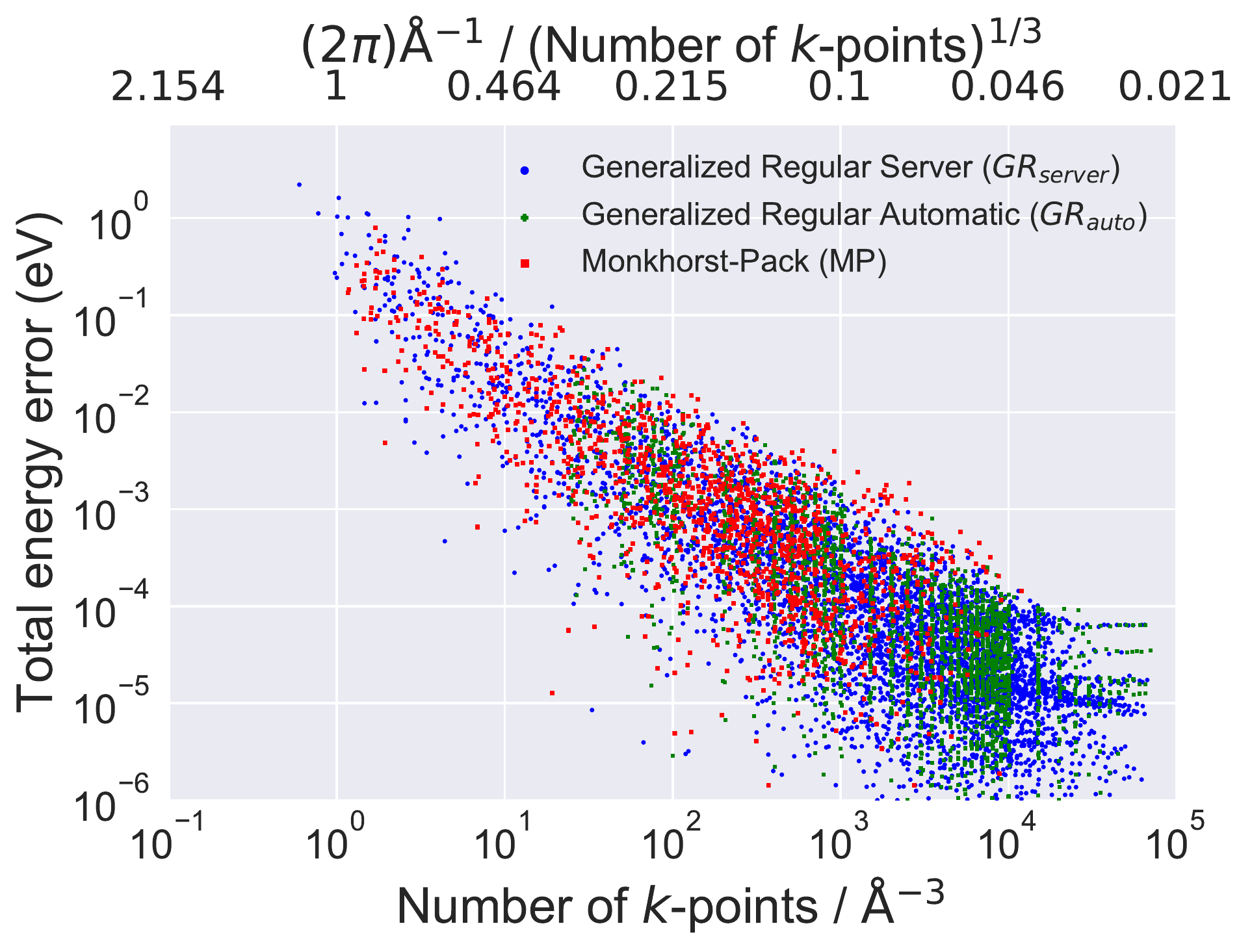}
  \caption{The total energy convergence with respect to total
    \kb-point density for MP, $GR_{\textrm{auto}}$ and
    $GR_{\textrm{server}}$ grids. The top axis shows the linear
    \kb-point spacing with a factor of 2$\pi$ included as part of the
    transformation to reciprocal space. This differs from the linear
    \kb-point spacing usually used as input in DFT codes by a factor
    of 2$\pi$, i.e., to get the spacing used as input in codes divide
    the values here by 2$\pi$.}
  \label{fig:total_scatter}
\end{figure}

\begin{figure}[t]
  \centering
  \includegraphics[width=8cm]{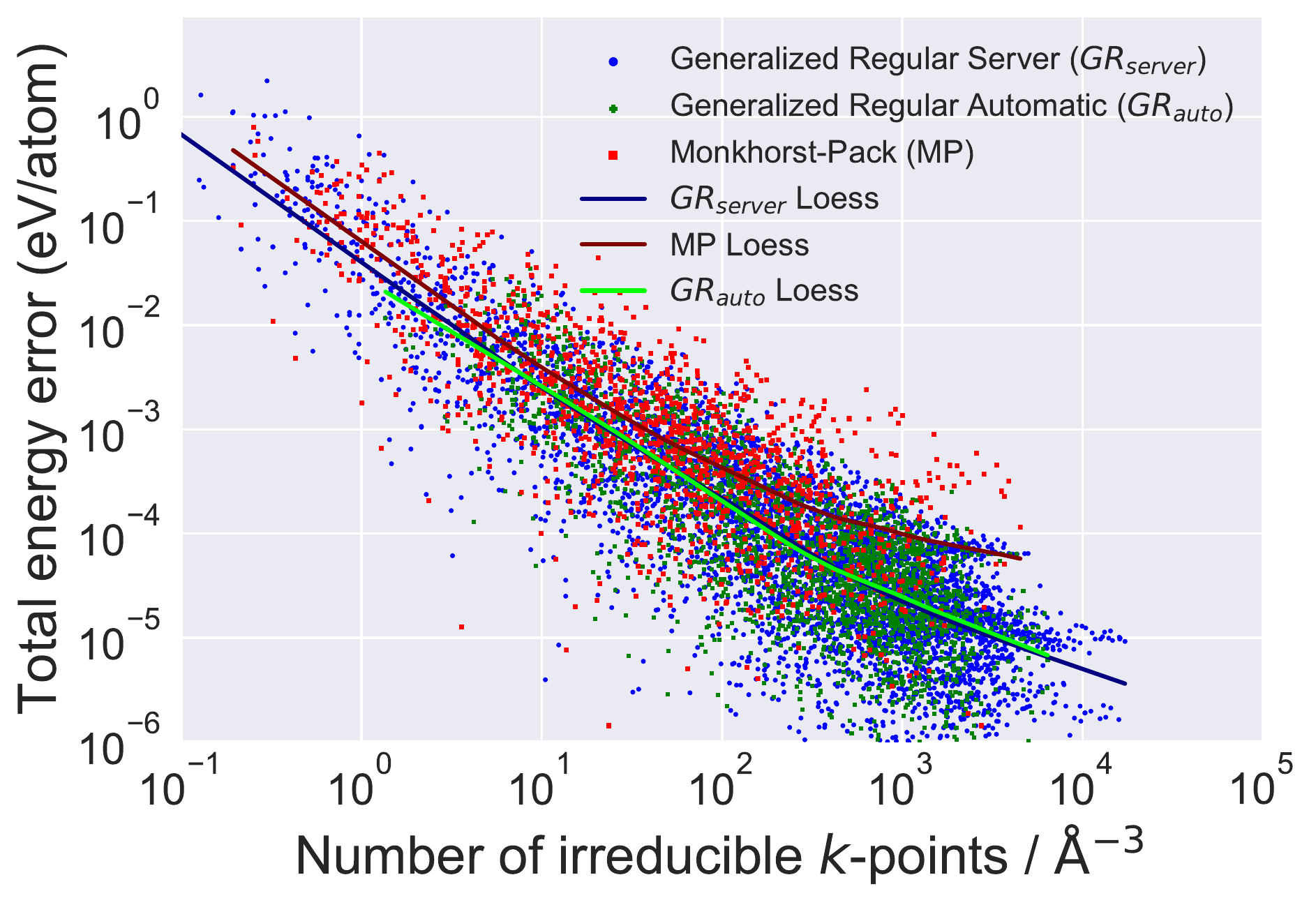}
  \caption{The total energy convergence with respect to irreducible
    \kb-point density for MP, $GR_{\textrm{auto}}$ and $GR_{\textrm{server}}$ grids with
    loess regression applied.}
  \label{fig:scatter}
\end{figure}

\begin{figure}[t]
  \centering
  \includegraphics[width=8cm]{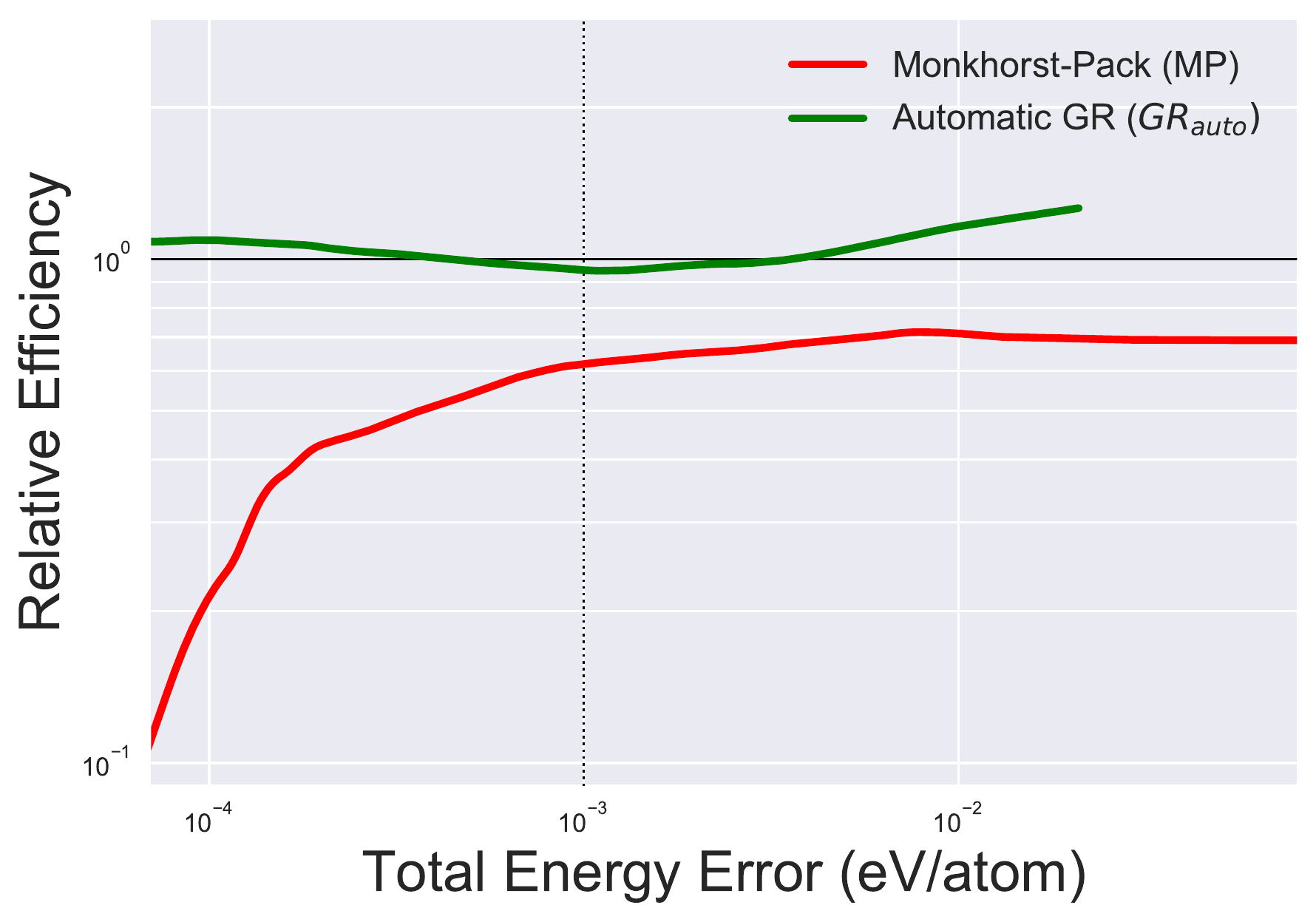}
  \caption{Along the $y$-axis are the ratios of the MP and
    $GR_{\textrm{auto}}$ efficiencies compared to the
    $GR_{\textrm{server}}$ grid efficiency (black horizontal line at
    $10^0$).  Total energy error (per atom) is plotted along the
    $x$-axis and decreases to the left.  MP grids are $\sim$60\% less
    efficient than both $GR_{\textrm{auto}}$ and
    $GR_{\textrm{server}}$ grids at a target accuracy of
    1meV/atom. The $GR_{\textrm{auto}}$ grids, however, outperform
    $GR_{\textrm{server}}$ grids at low densities but otherwise
    closely agree with $GR_{\textrm{auto}}$ grids.}
  \label{fig:loess}
\end{figure}

DFT calculations were performed using the Vienna Ab-initio Simulation
Package 4.6 (VASP 4.6) \cite{kresse1993ab, kresse1996efficiency,
  kresse1994ab, kresse1996efficient} on the nine monoatomic systems
mentioned above using PAW PBE
pseudopotentials.\cite{blochl1994projector, kresse1999ultrasoft} In
order to isolate the errors from \kb-point integeration, the different
cells were crystallographically equivalent to single element
cells. For MP grids, the target number of \kb-points ranged from
10-10,000 unreduced \kb-points, for $GR_{\textrm{server}}$ grids the
range was 4--240,000 unreduced \kb-points, and for
$GR_{\textrm{auto}}$ the range was 8 to 415,000 unreduced
\kb-points. In total, we compared errors across more than 7000 total
energy calculations. The energy taken as the error-free ``solution''
in our comparisons was the calculation with the highest \kb-point
density for each system. The total error convergence with respect to
the \kb-point density is shown in Fig.~\ref{fig:total_scatter}. The
total error convergence with repsect to the number of irreducible
\kb-points were compared using loess regression, see
Fig.~\ref{fig:scatter}. Ratios of these trend lines were then taken to
determine the efficiency of each grid relative to the
$GR_{\textrm{server}}$ grids (see Fig.~\ref{fig:loess}).

From Figs.~\ref{fig:scatter} and \ref{fig:loess}, it can be seen that
$GR_{\textrm{auto}}$ grids are up to $\sim$10\% more efficient and at
worst $\sim$5\% less efficient than $GR_{\textrm{server}}$ grids. Both
sets of grids outperform MP grids by $\sim$60\% at an accuracy target
of 1 meV/atom. The runtime for the algorithm to generate
$GR_{\textrm{auto}}$ grids at a \kb-point density of 5000 (dense
enough to achieve 1 meV/atom accuracy) was $\sim$3 seconds on average.

\section{Conclusion}

We have designed an algorithm that generates Generalized Regular (GR)
grids ``on the fly''. These $GR_{\textrm{auto}}$ grids are $\sim$60\%
more efficient than MP grids at an accuracy target of 1 meV/atom and
have have similar efficiency to $GR_{\textrm{server}}$
grids\cite{wisesa2016efficient}.

The algorithm is able to reduce the search space for $GR$ grids by
only generating grids that preserve the symmetry of the input
cell. The symmetry preserving grids are then filtered so that only the
most efficient grid is returned to the user. For our test cases the
average runtime of finding the optimal grid was $\sim$3 seconds. This
algorithm has been implemented and is available for download at:
https://github.com/msg-byu/GRkgridgen

\section{Acknowledgments}

The authors are grateful to Tim Mueller, Georg Kresse and Martijn
Marsman for helpful discussions. This work was supported by the Office
of Naval Research (ONR MURI N00014-13-1-0635). The authors are
grateful to C.S. Reese who helped with the loess regression and
statistical analysis of the data shown Figs.~\ref{fig:scatter} and
\ref{fig:loess}.

\section{Appendix}

  \subsection{Symmetry Preserving Offsets}

The following is a table of the symmetry preserving offsets for each
Bravais lattice expressed in terms of fractions of the primitive
lattice vectors.

{
\renewcommand{\arraystretch}{1.25}
\begin{table}[h]
\begin{tabular}{lll}
 Simple Cubic &  $\begin{pmatrix} \frac{1}{2}, \; \frac{1}{2}, \; \frac{1}{2} \end{pmatrix}$ \\
 Face Centered Cubic &  $\begin{pmatrix} \frac{1}{2}, \; \frac{1}{2}, \; \frac{1}{2} \end{pmatrix}$ \\
 Body Centered Cubic &  None \\
 Hexagonal &  $\begin{pmatrix} 0, \; 0, \; \frac{1}{2} \end{pmatrix}$ \\ 
 Rhombohedral &  $\begin{pmatrix} 0, \; 0, \; \frac{1}{2} \end{pmatrix}$ \\ 
 Simple Tetragonal &  $\begin{pmatrix} 0, \; 0, \; \frac{1}{2} \\ \frac{1}{2}, \; \frac{1}{2}, \; 0 \\ \frac{1}{2}, \; \frac{1}{2}, \; \frac{1}{2} \end{pmatrix}$ \\ 
 Body Centered Tetragonal & $\begin{pmatrix} 0, \; 0, \; \frac{1}{2} \end{pmatrix}$ \\ 
 Simple Orthorhombic &  $\begin{pmatrix} 0, \; 0, \; \frac{1}{2} \\ 0, \; \frac{1}{2}, \; 0 \\ \frac{1}{2}, \; 0, \; 0 \\ 0, \; \frac{1}{2}, \; \frac{1}{2} \\ \frac{1}{2}, \; 0, \; \frac{1}{2} \\ \frac{1}{2}, \; \frac{1}{2}, \; 0 \\ \frac{1}{2}, \; \frac{1}{2}, \; \frac{1}{2} \end{pmatrix}$ \\
 Base Centered Orthorhombic &  $\begin{pmatrix} 0, \; 0, \; \frac{1}{2} \\ 0, \; \frac{1}{2}, \; 0 \\ 0, \; \frac{1}{2}, \; \frac{1}{2} \end{pmatrix}$ \\
 Face Centered Orthorhombic &  $\begin{pmatrix} \frac{1}{2}, \; \frac{1}{2}, \; \frac{1}{2} \end{pmatrix}$ \\
 Body Centered Orthorhombic &  $\begin{pmatrix} 0, \; 0, \; \frac{1}{2} \\ 0, \; \frac{1}{2}, \; 0 \\ \frac{1}{2}, \; 0, \; 0 \end{pmatrix}$ \\
 Simple Monoclinic &  $\begin{pmatrix} 0, \; 0, \; \frac{1}{2} \\ 0, \; \frac{1}{2}, \; 0 \\ \frac{1}{2}, \; 0, \; 0 \\ 0, \; \frac{1}{2}, \; \frac{1}{2}, \\ \frac{1}{2} \; 0, \; \frac{1}{2}, \\ \frac{1}{2}, \; \frac{1}{2}, \; 0 \\ \frac{1}{2}, \; \frac{1}{2}, \; \frac{1}{2} \end{pmatrix}$ \\
 Base Centered Monoclinic & $\begin{pmatrix} 0, \; 0, \; \frac{1}{2} \\ 0, \; \frac{1}{2}, \; 0 \\ -\frac{1}{4}, \; \frac{1}{2}, \; 0 \\ -\frac{1}{4}, \; \frac{1}{4}, \; \frac{1}{2} \\ \frac{1}{4}, \; \frac{1}{4}, \; 0 \\ \frac{1}{4}, \; \frac{1}{4}, \; \frac{1}{2} \\ 0, \; \frac{1}{2}, \; \frac{1}{2} \end{pmatrix}$ \\
 Triclinic &  None
\end{tabular}
\end{table}
}





%


\end{document}